\documentclass[conference,10pt]{IEEEtran}
\IEEEoverridecommandlockouts

\usepackage{cite}
\usepackage[T1]{fontenc}
\usepackage{graphicx}
\usepackage{amssymb}
\usepackage{amsmath}
\usepackage{amsthm}
\usepackage{microtype}
\usepackage{balance}
\usepackage{subfigure}
\usepackage{xcolor} 
\usepackage{booktabs}

\usepackage{amsfonts}
\usepackage{algorithm}
\usepackage{algorithmic}
\usepackage{graphicx}
\usepackage{textcomp}
\usepackage{xcolor}
\usepackage{microtype}
\usepackage[caption=false,font=footnotesize]{subfig}
\usepackage{adjustbox}
\usepackage{balance}

\usepackage{tablefootnote}
\usepackage[caption=false,font=footnotesize]{subfig}

\usepackage{amssymb}
\usepackage{amsfonts}
\usepackage{mathrsfs}
\usepackage{xspace}
\usepackage{bm}
\usepackage{upgreek}

\newcommand{\safemath}[2]{\newcommand{#1}{\ensuremath{#2}\xspace}}

\safemath{\bma}{\mathbf{a}}
\safemath{\bmb}{\mathbf{b}}
\safemath{\bmc}{\mathbf{c}}
\safemath{\bmd}{\mathbf{d}}
\safemath{\bme}{\mathbf{e}}
\safemath{\bmf}{\mathbf{f}}
\safemath{\bmg}{\mathbf{g}}
\safemath{\bmh}{\mathbf{h}}
\safemath{\bmi}{\mathbf{i}}
\safemath{\bmj}{\mathbf{j}}
\safemath{\bmk}{\mathbf{k}}
\safemath{\bml}{\mathbf{l}}
\safemath{\bmm}{\mathbf{m}}
\safemath{\bmn}{\mathbf{n}}
\safemath{\bmo}{\mathbf{o}}
\safemath{\bmp}{\mathbf{p}}
\safemath{\bmq}{\mathbf{q}}
\safemath{\bmr}{\mathbf{r}}
\safemath{\bms}{\mathbf{s}}
\safemath{\bmt}{\mathbf{t}}
\safemath{\bmu}{\mathbf{u}}
\safemath{\bmv}{\mathbf{v}}
\safemath{\bmw}{\mathbf{w}}
\safemath{\bmx}{\mathbf{x}}
\safemath{\bmy}{\mathbf{y}}
\safemath{\bmz}{\mathbf{z}}
\safemath{\bmzero}{\mathbf{0}}
\safemath{\bmone}{\mathbf{1}}

\bmdefine{\biad}{a}
\bmdefine{\bibd}{b}
\bmdefine{\bicd}{c}
\bmdefine{\bidd}{d}
\bmdefine{\bied}{e}
\bmdefine{\bifd}{f}
\bmdefine{\bigd}{g}
\bmdefine{\bihd}{h}
\bmdefine{\biid}{i}
\bmdefine{\bijd}{j}
\bmdefine{\bikd}{k}
\bmdefine{\bild}{l}
\bmdefine{\bimd}{m}
\bmdefine{\bind}{n}
\bmdefine{\biod}{o}
\bmdefine{\bipd}{p}
\bmdefine{\biqd}{q}
\bmdefine{\bird}{r}
\bmdefine{\bisd}{s}
\bmdefine{\bitd}{t}
\bmdefine{\biud}{u}
\bmdefine{\bivd}{v}
\bmdefine{\biwd}{w}
\bmdefine{\bixd}{x}
\bmdefine{\biyd}{y}
\bmdefine{\bizd}{z}

\bmdefine{\bixid}{\xi}
\bmdefine{\bilambdad}{\lambda}
\bmdefine{\bimud}{\mu}
\bmdefine{\bithetad}{\theta}
\bmdefine{\biphid}{\phi}
\bmdefine{\bideltad}{\delta}

\safemath{\bmia}{\biad}
\safemath{\bmib}{\bibd}
\safemath{\bmic}{\bicd}
\safemath{\bmid}{\bidd}
\safemath{\bmie}{\bied}
\safemath{\bmif}{\bifd}
\safemath{\bmig}{\bigd}
\safemath{\bmih}{\bihd}
\safemath{\bmii}{\biid}
\safemath{\bmij}{\bijd}
\safemath{\bmik}{\bikd}
\safemath{\bmil}{\bild}
\safemath{\bmim}{\bimd}
\safemath{\bmin}{\bind}
\safemath{\bmio}{\biod}
\safemath{\bmip}{\bipd}
\safemath{\bmiq}{\biqd}
\safemath{\bmir}{\bird}
\safemath{\bmis}{\bisd}
\safemath{\bmit}{\bitd}
\safemath{\bmiu}{\biud}
\safemath{\bmiv}{\bivd}
\safemath{\bmiw}{\biwd}
\safemath{\bmix}{\bixd}
\safemath{\bmiy}{\biyd}
\safemath{\bmiz}{\bizd}

\safemath{\bmxi}{\bixid}
\safemath{\bmlambda}{\bilambdad}
\safemath{\bmmu}{\bimud}
\safemath{\bmtheta}{\bithetad}
\safemath{\bmphi}{\biphid}
\safemath{\bmdelta}{\bideltad}

\safemath{\bA}{\mathbf{A}}
\safemath{\bB}{\mathbf{B}}
\safemath{\bC}{\mathbf{C}}
\safemath{\bD}{\mathbf{D}}
\safemath{\bE}{\mathbf{E}}
\safemath{\bF}{\mathbf{F}}
\safemath{\bG}{\mathbf{G}}
\safemath{\bH}{\mathbf{H}}
\safemath{\bI}{\mathbf{I}}
\safemath{\bJ}{\mathbf{J}}
\safemath{\bK}{\mathbf{K}}
\safemath{\bL}{\mathbf{L}}
\safemath{\bM}{\mathbf{M}}
\safemath{\bN}{\mathbf{N}}
\safemath{\bO}{\mathbf{O}}
\safemath{\bP}{\mathbf{P}}
\safemath{\bQ}{\mathbf{Q}}
\safemath{\bR}{\mathbf{R}}
\safemath{\bS}{\mathbf{S}}
\safemath{\bT}{\mathbf{T}}
\safemath{\bU}{\mathbf{U}}
\safemath{\bV}{\mathbf{V}}
\safemath{\bW}{\mathbf{W}}
\safemath{\bX}{\mathbf{X}}
\safemath{\bY}{\mathbf{Y}}
\safemath{\bZ}{\mathbf{Z}}

\safemath{\bZero}{\mathbf{0}}
\safemath{\bOne}{\mathbf{1}}
\safemath{\bDelta}{\mathbf{\Delta}}
\safemath{\bLambda}{\mathbf{\UpLambda}}
\safemath{\bPhi}{\mathbf{\Upphi}}
\safemath{\bSigma}{\mathbf{\Upsigma}}
\safemath{\bOmega}{\mathbf{\Upomega}}
\safemath{\bTheta}{\mathbf{\Uptheta}}

\bmdefine{\biAd}{A}
\bmdefine{\biBd}{B}
\bmdefine{\biCd}{C}
\bmdefine{\biDd}{D}
\bmdefine{\biEd}{E}
\bmdefine{\biFd}{F}
\bmdefine{\biGd}{G}
\bmdefine{\biHd}{H}
\bmdefine{\biId}{I}
\bmdefine{\biJd}{J}
\bmdefine{\biKd}{K}
\bmdefine{\biLd}{L}
\bmdefine{\biMd}{M}
\bmdefine{\biOd}{N}
\bmdefine{\biPd}{O}
\bmdefine{\biQd}{P}
\bmdefine{\biRd}{R}
\bmdefine{\biSd}{S}
\bmdefine{\biTd}{T}
\bmdefine{\biUd}{U}
\bmdefine{\biVd}{V}
\bmdefine{\biWd}{W}
\bmdefine{\biXd}{X}
\bmdefine{\biYd}{Y}
\bmdefine{\biZd}{Z}

\bmdefine{\biDelta}{\Delta}
\bmdefine{\biLambda}{\Lambda}
\bmdefine{\biPhi}{\Phi}
\bmdefine{\biSigma}{\Sigma}
\bmdefine{\biOmega}{\Omega}
\bmdefine{\biTheta}{\Theta}

\safemath{\bimA}{\biAd}
\safemath{\bimB}{\biBd}
\safemath{\bimC}{\biCd}
\safemath{\bimD}{\biDd}
\safemath{\bimE}{\biEd}
\safemath{\bimF}{\biFd}
\safemath{\bimG}{\biGd}
\safemath{\bimH}{\biHd}
\safemath{\bimI}{\biId}
\safemath{\bimJ}{\biJd}
\safemath{\bimK}{\biKd}
\safemath{\bimL}{\biLd}
\safemath{\bimM}{\biMd}
\safemath{\bimN}{\biNd}
\safemath{\bimO}{\biOd}
\safemath{\bimP}{\biPd}
\safemath{\bimQ}{\biQd}
\safemath{\bimR}{\biRd}
\safemath{\bimS}{\biSd}
\safemath{\bimT}{\biTd}
\safemath{\bimU}{\biUd}
\safemath{\bimV}{\biVd}
\safemath{\bimW}{\biWd}
\safemath{\bimX}{\biXd}
\safemath{\bimY}{\biYd}
\safemath{\bimZ}{\biZd}

\safemath{\bimDelta}{\biDelta}
\safemath{\bimLambda}{\biLambda}
\safemath{\bimPhi}{\biPhi}
\safemath{\bimSigma}{\biSigma}
\safemath{\bimOmega}{\biOmega}
\safemath{\bimTheta}{\biTheta}

\safemath{\setA}{\mathcal{A}}
\safemath{\setB}{\mathcal{B}}
\safemath{\setC}{\mathcal{C}}
\safemath{\setD}{\mathcal{D}}
\safemath{\setE}{\mathcal{E}}
\safemath{\setF}{\mathcal{F}}
\safemath{\setG}{\mathcal{G}}
\safemath{\setH}{\mathcal{H}}
\safemath{\setI}{\mathcal{I}}
\safemath{\setJ}{\mathcal{J}}
\safemath{\setK}{\mathcal{K}}
\safemath{\setL}{\mathcal{L}}
\safemath{\setM}{\mathcal{M}}
\safemath{\setN}{\mathcal{N}}
\safemath{\setO}{\mathcal{O}}
\safemath{\setP}{\mathcal{P}}
\safemath{\setQ}{\mathcal{Q}}
\safemath{\setR}{\mathcal{R}}
\safemath{\setS}{\mathcal{S}}
\safemath{\setT}{\mathcal{T}}
\safemath{\setU}{\mathcal{U}}
\safemath{\setV}{\mathcal{V}}
\safemath{\setW}{\mathcal{W}}
\safemath{\setX}{\mathcal{X}}
\safemath{\setY}{\mathcal{Y}}
\safemath{\setZ}{\mathcal{Z}}
\safemath{\emptySet}{\varnothing}

\safemath{\colA}{\mathscr{A}}
\safemath{\colB}{\mathscr{B}}
\safemath{\colC}{\mathscr{C}}
\safemath{\colD}{\mathscr{D}}
\safemath{\colE}{\mathscr{E}}
\safemath{\colF}{\mathscr{F}}
\safemath{\colG}{\mathscr{G}}
\safemath{\colH}{\mathscr{H}}
\safemath{\colI}{\mathscr{I}}
\safemath{\colJ}{\mathscr{J}}
\safemath{\colK}{\mathscr{K}}
\safemath{\colL}{\mathscr{L}}
\safemath{\colM}{\mathscr{M}}
\safemath{\colN}{\mathscr{N}}
\safemath{\colO}{\mathscr{O}}
\safemath{\colP}{\mathscr{P}}
\safemath{\colQ}{\mathscr{Q}}
\safemath{\colR}{\mathscr{R}}
\safemath{\colS}{\mathscr{S}}
\safemath{\colT}{\mathscr{T}}
\safemath{\colU}{\mathscr{U}}
\safemath{\colV}{\mathscr{V}}
\safemath{\colW}{\mathscr{W}}
\safemath{\colX}{\mathscr{X}}
\safemath{\colY}{\mathscr{Y}}
\safemath{\colZ}{\mathscr{Z}}

\safemath{\opA}{\mathbb{A}}
\safemath{\opB}{\mathbb{B}}
\safemath{\opC}{\mathbb{C}}
\safemath{\opD}{\mathbb{D}}
\safemath{\opE}{\mathbb{E}}
\safemath{\opF}{\mathbb{F}}
\safemath{\opG}{\mathbb{G}}
\safemath{\opH}{\mathbb{H}}
\safemath{\opI}{\mathbb{I}}
\safemath{\opJ}{\mathbb{J}}
\safemath{\opK}{\mathbb{K}}
\safemath{\opL}{\mathbb{L}}
\safemath{\opM}{\mathbb{M}}
\safemath{\opN}{\mathbb{N}}
\safemath{\opO}{\mathbb{O}}
\safemath{\opP}{\mathbb{P}}
\safemath{\opQ}{\mathbb{Q}}
\safemath{\opR}{\mathbb{R}}
\safemath{\opS}{\mathbb{S}}
\safemath{\opT}{\mathbb{T}}
\safemath{\opU}{\mathbb{U}}
\safemath{\opV}{\mathbb{V}}
\safemath{\opW}{\mathbb{W}}
\safemath{\opX}{\mathbb{X}}
\safemath{\opY}{\mathbb{Y}}
\safemath{\opZ}{\mathbb{Z}}
\safemath{\opZero}{\mathbb{O}}
\safemath{\identityop}{\opI}

\safemath{\veca}{\bma}
\safemath{\vecb}{\bmb}
\safemath{\vecc}{\bmc}
\safemath{\vecd}{\bmd}
\safemath{\vece}{\bme}
\safemath{\vecf}{\bmf}
\safemath{\vecg}{\bmg}
\safemath{\vech}{\bmh}
\safemath{\veci}{\bmi}
\safemath{\vecj}{\bmj}
\safemath{\veck}{\bmk}
\safemath{\vecl}{\bml}
\safemath{\vecm}{\bmm}
\safemath{\vecn}{\bmn}
\safemath{\veco}{\bmo}
\safemath{\vecp}{\bmp}
\safemath{\vecq}{\bmq}
\safemath{\vecr}{\bmr}
\safemath{\vecs}{\bms}
\safemath{\vect}{\bmt}
\safemath{\vecu}{\bmu}
\safemath{\vecv}{\bmv}
\safemath{\vecw}{\bmw}
\safemath{\vecx}{\bmx}
\safemath{\vecy}{\bmy}
\safemath{\vecz}{\bmz}

\safemath{\veczero}{\bmzero}
\safemath{\vecone}{\bmone}
\safemath{\vecxi}{\bmxi}
\safemath{\veclambda}{\bmlambda}
\safemath{\vecmu}{\bmmu}
\safemath{\vectheta}{\bmtheta}
\safemath{\vecphi}{\bmphi}
\safemath{\vecdelta}{\bmdelta}

\safemath{\matA}{\bA}
\safemath{\matB}{\bB}
\safemath{\matC}{\bC}
\safemath{\matD}{\bD}
\safemath{\matE}{\bE}
\safemath{\matF}{\bF}
\safemath{\matG}{\bG}
\safemath{\matH}{\bH}
\safemath{\matI}{\bI}
\safemath{\matJ}{\bJ}
\safemath{\matK}{\bK}
\safemath{\matL}{\bL}
\safemath{\matM}{\bM}
\safemath{\matN}{\bN}
\safemath{\matO}{\bO}
\safemath{\matP}{\bP}
\safemath{\matQ}{\bQ}
\safemath{\matR}{\bR}
\safemath{\matS}{\bS}
\safemath{\matT}{\bT}
\safemath{\matU}{\bU}
\safemath{\matV}{\bV}
\safemath{\matW}{\bW}
\safemath{\matX}{\bX}
\safemath{\matY}{\bY}
\safemath{\matZ}{\bZ}
\safemath{\matzero}{\bmzero}

\safemath{\matDelta}{\bDelta}
\safemath{\matLambda}{\bLambda}
\safemath{\matPhi}{\bPhi}
\safemath{\matSigma}{\bSigma}
\safemath{\matOmega}{\bOmega}
\safemath{\matTheta}{\bTheta}

\safemath{\matidentity}{\matI}
\safemath{\matone}{\matO}

\safemath{\rnda}{A}
\safemath{\rndb}{B}
\safemath{\rndc}{C}
\safemath{\rndd}{D}
\safemath{\rnde}{E}
\safemath{\rndf}{F}
\safemath{\rndg}{G}
\safemath{\rndh}{H}
\safemath{\rndi}{I}
\safemath{\rndj}{J}
\safemath{\rndk}{K}
\safemath{\rndl}{L}
\safemath{\rndm}{M}
\safemath{\rndn}{N}
\safemath{\rndo}{O}
\safemath{\rndp}{P}
\safemath{\rndq}{Q}
\safemath{\rndr}{R}
\safemath{\rnds}{S}
\safemath{\rndt}{T}
\safemath{\rndu}{U}
\safemath{\rndv}{V}
\safemath{\rndw}{W}
\safemath{\rndx}{X}
\safemath{\rndy}{Y}
\safemath{\rndz}{Z}

\safemath{\rveca}{\bimA}
\safemath{\rvecb}{\bimB}
\safemath{\rvecc}{\bimC}
\safemath{\rvecd}{\bimD}
\safemath{\rvece}{\bimE}
\safemath{\rvecf}{\bimF}
\safemath{\rvecg}{\bimG}
\safemath{\rvech}{\bimH}
\safemath{\rveci}{\bimI}
\safemath{\rvecj}{\bimJ}
\safemath{\rveck}{\bimK}
\safemath{\rvecl}{\bimL}
\safemath{\rvecm}{\bimM}
\safemath{\rvecn}{\bimN}
\safemath{\rveco}{\bomO}
\safemath{\rvecp}{\bimP}
\safemath{\rvecq}{\bimQ}
\safemath{\rvecr}{\bimR}
\safemath{\rvecs}{\bimS}
\safemath{\rvect}{\bimT}
\safemath{\rvecu}{\bimU}
\safemath{\rvecv}{\bimV}
\safemath{\rvecw}{\bimW}
\safemath{\rvecx}{\bimX}
\safemath{\rvecy}{\bimY}
\safemath{\rvecz}{\bimZ}

\safemath{\rvecxi}{\bmxi}
\safemath{\rveclambda}{\bmlambda}
\safemath{\rvecmu}{\bmmu}
\safemath{\rvectheta}{\bmtheta}
\safemath{\rvecphi}{\bmphi}

\safemath{\rmatA}{\bimA}
\safemath{\rmatB}{\bimB}
\safemath{\rmatC}{\bimC}
\safemath{\rmatD}{\bimD}
\safemath{\rmatE}{\bimE}
\safemath{\rmatF}{\bimF}
\safemath{\rmatG}{\bimG}
\safemath{\rmatH}{\bimH}
\safemath{\rmatI}{\bimI}
\safemath{\rmatJ}{\bimJ}
\safemath{\rmatK}{\bimK}
\safemath{\rmatL}{\bimL}
\safemath{\rmatM}{\bimM}
\safemath{\rmatN}{\bimN}
\safemath{\rmatO}{\bimO}
\safemath{\rmatP}{\bimP}
\safemath{\rmatQ}{\bimQ}
\safemath{\rmatR}{\bimR}
\safemath{\rmatS}{\bimS}
\safemath{\rmatT}{\bimT}
\safemath{\rmatU}{\bimU}
\safemath{\rmatV}{\bimV}
\safemath{\rmatW}{\bimW}
\safemath{\rmatX}{\bimX}
\safemath{\rmatY}{\bimY}
\safemath{\rmatZ}{\bimZ}

\safemath{\rmatDelta}{\bimDelta}
\safemath{\rmatLambda}{\bimLambda}
\safemath{\rmatPhi}{\bimPhi}
\safemath{\rmatSigma}{\bimSigma}
\safemath{\rmatOmega}{\bimOmega}
\safemath{\rmatTheta}{\bimTheta}

\usepackage{amssymb}
\usepackage{amsfonts}
\usepackage{mathrsfs}
\usepackage{xspace}
\usepackage{bm}
\usepackage{fancyref}
\usepackage{textcomp}

\usepackage{multirow}
\usepackage{stmaryrd}

\newenvironment{textbmatrix}{	\setlength{\arraycolsep}{2.5pt}%
								\big[\begin{matrix}}{\end{matrix}\big]%
								\raisebox{0.08ex}{\vphantom{M}}}

\def\be{\begin{equation}}
\def\ee{\end{equation}}
\def\een{\nonumber \end{equation}}
\def\mat{\begin{bmatrix}}
\def\emat{\end{bmatrix}}
\def\btm{\begin{textbmatrix}}
\def\etm{\end{textbmatrix}}

\def\ba#1\ea{\begin{align}#1\end{align}}
\def\bas#1\eas{\begin{align*}#1\end{align*}}
\def\bs#1\es{\begin{split}#1\end{split}}
\def\bg#1\eg{\begin{gather}#1\end{gather}}
\def\bml#1\eml{\begin{multline}#1\end{multline}}
\def\bi#1\ei{\begin{itemize}#1\end{itemize}}

\safemath{\dirac}{\delta}					
\safemath{\krond}{\dirac}					

\safemath{\upto}{\uparrow}
\safemath{\downto}{\downarrow}
\safemath{\iu}{j}							
\safemath{\ev}{\lambda}						
\safemath{\hilseqspace}{l^{2}}				
\newcommand{\banachfunspace}[1]{\setL^{#1}}	
\safemath{\hilfunspace}{\banachfunspace{2}}	

\safemath{\SNR}{\textit{SNR}} 				
\safemath{\PAR}{\textit{PAR}} 				
\safemath{\No}{N_0}							
\safemath{\Es}{E_s}							
\safemath{\Eb}{E_b}							
\safemath{\EbNo}{\frac{\Eb}{\No}}
\safemath{\EsNo}{\frac{\Es}{\No}}

\DeclareMathOperator{\CHop}{\ensuremath{\opH}} 
\safemath{\tvir}{\rndh_{\CHop}}				
\safemath{\tvtf}{\rndl_{\CHop}}				
\safemath{\spf}{\rnds_{\CHop}}				
\safemath{\bff}{H_{\CHop}}					

\safemath{\ircf}{r_{h}}						
\safemath{\tftvcf}{r_{s}}					
\safemath{\tfcf}{r_{l}}						
\safemath{\bfcf}{r_{H}}						

\safemath{\tcorr}{c_h}						
\safemath{\scf}{c_{s}}						
\safemath{\tfcorr}{c_{l}}					
\safemath{\fcorr}{c_{H}}						

\safemath{\mi}{I}							
\safemath{\capacity}{C}						

\safemath{\normal}{\mathcal{N}}			
\safemath{\jpg}{\mathcal{CN}}			
\safemath{\mchain}{\leftrightarrow}		

\safemath{\dB}{\,\mathrm{dB}}
\safemath{\dBm}{\,\mathrm{dBm}}
\safemath{\Hz}{\,\mathrm{Hz}}
\safemath{\kHz}{\,\mathrm{kHz}}
\safemath{\MHz}{\,\mathrm{MHz}}
\safemath{\GHz}{\,\mathrm{GHz}}
\safemath{\s}{\,\mathrm{s}}
\safemath{\ms}{\,\mathrm{ms}}
\safemath{\mus}{\,\mathrm{\text{\textmu}s}}
\safemath{\ns}{\,\mathrm{ns}}
\safemath{\ps}{\,\mathrm{ps}}
\safemath{\meter}{\,\mathrm{m}}
\safemath{\mm}{\,\mathrm{mm}}
\safemath{\cm}{\,\mathrm{cm}}
\safemath{\m}{\,\mathrm{m}}
\safemath{\W}{\,\mathrm{W}}
\safemath{\mW}{\, \mathrm{mW}}
\safemath{\J}{\,\mathrm{J}}
\safemath{\K}{\,\mathrm{K}}
\safemath{\bit}{\,\mathrm{bit}}
\safemath{\nat}{\,\mathrm{nat}}

\safemath{\define}{\triangleq}			

\safemath{\equivalent}{\sim}
\safemath{\distas}{\sim}					
\safemath{\sdiff}{\Delta}				

\safemath{\reals}{\mathbb{R}}
\safemath{\positivereals}{\reals_{+}}
\safemath{\integers}{\mathbb{Z}}
\safemath{\posint}{\integers_{+}}
\safemath{\naturals}{\mathbb{N}}
\safemath{\posnaturals}{\naturals_{+}}
\safemath{\complexset}{\mathbb{C}}
\safemath{\rationals}{\mathbb{Q}}

\newcommand*{\fancyrefapplabelprefix}{app}		
\newcommand*{\fancyrefthmlabelprefix}{thm}		
\newcommand*{\fancyreflemlabelprefix}{lem}		
\newcommand*{\fancyrefcorlabelprefix}{cor}		
\newcommand*{\fancyrefdeflabelprefix}{def}		
\newcommand*{\fancyrefproplabelprefix}{prop}		
\newcommand*{\fancyrefexmpllabelprefix}{exmpl}
\newcommand*{\fancyrefalglabelprefix}{alg}		
\newcommand*{\fancyreftbllabelprefix}{tbl}		

\frefformat{vario}{\fancyrefseclabelprefix}{Sec.~#1}
\frefformat{vario}{\fancyrefthmlabelprefix}{Thm.~#1}
\frefformat{vario}{\fancyreftbllabelprefix}{Tbl.~#1}
\frefformat{vario}{\fancyreflemlabelprefix}{Lem.~#1}
\frefformat{vario}{\fancyrefcorlabelprefix}{Cor.~#1}
\frefformat{vario}{\fancyrefdeflabelprefix}{Def.~#1}
\frefformat{vario}{\fancyreffiglabelprefix}{Fig.~#1}
\frefformat{vario}{\fancyrefapplabelprefix}{App.~#1}
\frefformat{vario}{\fancyrefeqlabelprefix}{(#1)}
\frefformat{vario}{\fancyrefproplabelprefix}{Prop.~#1}
\frefformat{vario}{\fancyrefexmpllabelprefix}{Ex.~#1}
\frefformat{vario}{\fancyrefalglabelprefix}{Alg.~#1}

\safemath{\dictab}{[\,\dicta\,\,\dictb\,]}

\safemath{\ysig}{\bmy}
\safemath{\ysighat}{\hat{\ysig}}
\safemath{\ysigdim}{M}
\safemath{\xsig}{\bmx}
\safemath{\xsigdim}{N}
\safemath{\nx}{n_x}
\safemath{\zsig}{\bmz}
\safemath{\zsigdim}{\ysigdim}
\safemath{\rsig}{\bmr}
\safemath{\Adict}{\bA}
\safemath{\Adicttilde}{\widetilde{\Adict}}
\safemath{\Adictdim}{\outputdim\times\xsigdim}
\safemath{\avec}{\bma}
\safemath{\avectilde}{\tilde{\avec}}
\safemath{\Bdict}{\bB}
\safemath{\Bdicttilde}{\widetilde{\Bdict}}
\safemath{\Cdict}{\bC}
\safemath{\cvec}{\bmc}
\safemath{\Ddict}{\bD}
\safemath{\Ddictdim}{\ysigdim\times\xsigdim}
\safemath{\dvec}{\bmd}
\safemath{\Ddicttilde}{\widetilde{\bD}}
\safemath{\Bonb}{\bB}
\safemath{\bvec}{\bmb}
\safemath{\Bonbdim}{\ysigdim\times\ysigdim}
\safemath{\noise}{\bmn}
\safemath{\noisedim}{\ysigim}
\safemath{\err}{\bme}
\safemath{\errdim}{\ysigdim}
\safemath{\errset}{\setE}
\safemath{\nerr}{n_e}
\safemath{\delop}{\bP_\errset}
\safemath{\delopc}{\bP_{{\errset}^c}}

\safemath{\cplxi}{\imath}
\safemath{\cplxj}{\jmath}

\safemath{\dict}{\matD}
\safemath{\inputdim}{N}		
\safemath{\outputdim}{M}		
\safemath{\sparsity}{S}	
\safemath{\inputdimA}{{N_a}}	
\safemath{\inputdimB}{{N_b}}	
\safemath{\elemA}{{n_a}}	
\safemath{\elemB}{{n_b}}	
\safemath{\resA}{\matR_a}	
\safemath{\resB}{\matR_b}	
\safemath{\subD}{\matS} 
\safemath{\subA}{\matS_a} 
\safemath{\subB}{\matS_b} 
\safemath{\dicta}{\matA} 	
\safemath{\dictb}{\matB} 	
\safemath{\hollowS}{H}
\safemath{\hollowA}{H_a}
\safemath{\hollowB}{H_b}
\safemath{\cross}{Z}
\safemath{\coh}{\mu_d}			
\safemath{\coha}{\mu_a}			
\safemath{\cohb}{\mu_b}			
\safemath{\mubs}{\nu}	
\safemath{\cohm}{\mu_m} 
\safemath{\dictset}{\setD}	
\safemath{\dictsetp}{\dictset(\coh,\coha,\cohb)}	
\safemath{\dictsetgen}{\dictset_\text{gen}}
\safemath{\dictsetgenp}{\dictsetgen(\coh)}
\safemath{\dictsetonb}{\dictset_\text{onb}}
\safemath{\dictsetonbp}{\dictsetonb(\coh)}

\safemath{\leftside}{U}
\safemath{\rightsideA}{R_a}
\safemath{\rightsideB}{R_b}

\safemath{\indexS}{\setI_S} 

\safemath{\na}{n_a}			
\safemath{\nb}{n_b}			
\safemath{\coeffa}{p_i}	
\safemath{\coeffb}{q_j}	
\safemath{\seta}{\setP}		
\safemath{\setb}{\setQ}     
\safemath{\setw}{\setW}	
\safemath{\setz}{\setZ}	
\safemath{\cola}{\veca}		
\safemath{\colb}{\vecb}		
\safemath{\cold}{\vecd}		
\safemath{\inputvec}{\vecx} 	
\safemath{\error}{\vece}	
\safemath{\noiseout}{\vecz} 	
\safemath{\inputvecel}{x}
\safemath{\inputveca}{\vecx_a}
\safemath{\inputvecb}{\vecx_b}
\safemath{\outputvec}{\vecy}	
\safemath{\lambdamin}{\lambda_{\mathrm{min}}}

\safemath{\elltwo}{\ell_2}
\safemath{\ellone}{\ell_1}
\safemath{\ellzero}{\ell_0}
\safemath{\ellinf}{\ell_\infty}
\safemath{\ellinftilde}{\ell_{\widetilde\infty}}
\safemath{\licard}{Z(\coh,\coha,\cohb)}
\safemath{\xsol}{\hat{x}}
\safemath{\xbord}{x_b}		
\safemath{\xstat}{x_s}		
\safemath{\xstatLone}{\tilde{x}_s}
\safemath{\order}{\mathcal{O}} 
\safemath{\scales}{\Theta} 
\safemath{\ones}{\mathbf{1}} 
\safemath{\zeroes}{\mathbf{0}} 
\safemath{\thlone}{\kappa(\coh,\cohb)} 
\safemath{\constoneA}{\delta} 
\safemath{\constoneB}{\epsilon} 
\safemath{\nlarge}{L}				   
\safemath{\sumlarge}{S_\nlarge}
\safemath{\maxlarger}{P_\nlarge}	   
\safemath{\Pzero}{\textrm{P0}}	
\safemath{\Pone}{\textrm{P1}}
\safemath{\vecfir}{\vecw}			 
\safemath{\vecsec}{\vecz}
\safemath{\elvecfir}{w}              
\safemath{\elvecsec}{z}				 
\safemath{\nlargefir}{n}
\safemath{\normout}{\gamma}
\safemath{\auxfun}{h}
\safemath{\supp}{\textrm{supp}}

\safemath{\indexa}{\ell}
\safemath{\indexb}{r}
\safemath{\indexc}{i}
\safemath{\indexd}{j}

\safemath{\project}{P}

\newcommand*{\fancyrefremarklabelprefix}{remark}
\frefformat{vario}{\fancyrefremarklabelprefix}{Remark~#1}

\usepackage{color}
\def\MSE{\mathrm{MSE}}
\newcommand{\xap}[1]{\text{\underbar{$\vecx$}}^{(#1)}}
\newcommand{\tildexap}[1]{\text{\underbar{$\tilde\vecx$}}^{(#1)}}
\def\loss{\mathfrak{L}}

\IEEEoverridecommandlockouts 
\allowdisplaybreaks 


\title{Channel Charting in Real-World Coordinates
\author{\IEEEauthorblockN{Sueda Taner, Victoria Palhares, and Christoph Studer} \\[0.05cm]
\thanks{The work of ST, VP, and CS was supported in part by the Swiss National Science Foundation (SNSF)
grant 207314 and an ETH Research grant. The work of CS was also supported in part by the US National Science Foundation (NSF) under grants CNS-1717559 and ECCS-1824379.}
\thanks{The authors would like to thank Paul Ferrand, Maxime Guillaud, and Olav Tirkkonen for discussions on channel charting {and Gian Marti for comments on the final version of this paper}. The authors also thank Remcom for providing a license for Wireless InSite \cite{remcom}.}
\IEEEauthorblockA{\em Department of Information Technology and Electrical Engineering, ETH Zurich, Switzerland \\
email: taners@iis.ee.ethz.ch, vmenescal@ethz.ch, and studer@ethz.ch} 
}}

\begin{document}

\maketitle


\begin{abstract}
Channel charting is an emerging self-supervised method that maps channel state information (CSI) to a low-dimensional latent space, which represents pseudo-positions of user equipments (UEs).
While this latent space preserves local geometry, i.e., nearby UEs are nearby in latent space, the pseudo-positions are in arbitrary coordinates and global geometry is not preserved.
In order to enable channel charting in real-world coordinates, we propose a novel bilateration loss for multipoint wireless systems in which only the access point (AP) locations are known---no geometrical models or ground-truth UE position information is required. 
The idea behind this bilateration loss is to compare the received power at pairs of APs in order to determine whether a UE should be placed closer to one AP or the other in latent space. 
We demonstrate the efficacy of our method using channel vectors from a commercial ray-tracer.
\end{abstract}


\section{Introduction} 
\label{sec:intro}

Channel charting is a self-supervised method that extracts user equipment (UE) pseudo-position solely by processing estimated channel state information (CSI) at infrastructure basestations (BSs) or access points (APs)~\cite{studer18cc}. 
The key idea is to apply dimensionality reduction to a large database of CSI features that represent large-scale properties of the wireless channel; this leads to a low-dimensional latent space---the so-called channel chart---that is tied to UE position.
This UE pseudo-position information can be used to assist a wide range of applications in wireless communication systems, such as pilot allocation~\cite{ribeiro22ojcs}, beam management\mbox{\cite{luclemag22,kazemi22vtc}}, channel capacity prediction \cite{kallehague23}, and many more~\cite{ferrand2023wireless}.

Although channel charts preserve the local geometry of UE positions, i.e., nearby UEs correspond to points that are nearby in the channel chart, global geometry is typically distorted and the learned channel charts are represented in arbitrary coordinates---all of this is a consequence of self-supervised learning. 
Multipoint channel charting techniques, which process CSI acquired at multiple distributed BSs or APs, have been shown to improve global geometry\mbox{\cite{deng18,euchner22,agostini22federated}}. Nonetheless, the resulting channel charts remain to be pseudo-positions with no direct ties to real-world coordinates.

A channel chart in real-world coordinates would improve spatial awareness, e.g., for handover purposes, and allow for more effective deployment planning. 
Therefore,  methods that associate channel charts with real-world coordinates have been proposed recently. 
One strain of methods first employs self-supervised multipoint channel charting and then, maps the resulting pseudo-positions to real-world coordinates~\cite{pihljasalo20,stahlke23}. 
The method in \cite{pihljasalo20} relies on the assumption that the received power is the highest for UE positions nearest to the AP, which may not hold true as we will demonstrate in \fref{sec:power_vs_distance}; the method in \cite{stahlke23} relies on accurately synchronized APs to incorporate time-of-flight-based ranging.
Another strain of methods deploys semi-supervised learning to anchor the channel chart in real-world coordinates~\cite{penghzipaper,lei19siamese,deng21networkside,karmanov21wicluster,zhang21globecom,deng21tsne}; all of these methods require ground-truth UE position information during training.

\subsection{Contributions} 

In contrast to prior work, we propose a novel method that learns neural network-based channel charting functions that generate channel charts in real-world coordinate systems.
The proposed method is weakly-supervised as it only exploits knowledge of the AP locations in a multipoint scenario---\emph{no} geometric models, accurate AP synchronization, or labeled CSI samples with UE ground-truth positions are required. 
More specifically, we postulate a novel bilateration loss, which compares the power of the received CSI between a transmitting UE and a pair of APs with the goal of placing the UE closer to the AP that is receiving higher power. 
While this bilateration loss alone already produces utilizable channel charts in real-world coordinates, their quality can be improved substantially when combined with the timestamp-based triplet loss from~\cite{ferrand2021}.
We demonstrate the efficacy of our method using channel vectors from a commercial ray-tracer \cite{remcom}.

\subsection{Notation}

Column vectors and matrices are denoted by lowercase and uppercase boldface letters, respectively; sets are denoted by uppercase calligraphic letters.
For a vector~$\bma$, 
the Euclidean norm is $\|\bma\|$ and the entry-wise absolute value is $|\bma|$. 
For a matrix $\bA$, the Frobenius norm is $\|\bA\|_F$ and the column-wise vectorization is $\mathrm{vec}(\bA)$. 
The cardinality of a set $\setA$ is $|\setA|$.
The operator $(x)^+=\max\{x,0\}$ is the rectified linear unit (ReLU).
%


\section{Channel Charting Basics}
\label{sec:background}

We now briefly outline the basics of channel charting, detail the system model, and discuss channel charting with neural networks using the triplet-based learning approach, which is part of the method we propose in \fref{sec:ccinreal}.

\subsection{Operating Principle}
Channel charting typically operates in two phases~\cite{studer18cc}. In the first phase, CSI from a large number of different UE positions is acquired and CSI features that capture large-scale fading properties of the channel are stored in a database. By applying parametric dimensionality reduction \cite{vandermaaten2009dimensionality} to the CSI feature database, one then learns a channel charting function (e.g., using a neural network), which maps CSI features to a low-dimensional representation: the channel chart.
Channel charts have the useful property that the UEs transmitting from nearby positions are also placed nearby in the channel chart. 
We reiterate that channel chart learning is self-supervised, meaning that no ground-truth information about the UEs' position is required.
In the second phase, the channel charting function is used to map new CSI features to points in the channel chart, which represent the transmitting UEs' pseudo-positions. 

\subsection{System Model}
\label{sec:system_model}

We consider a {single-input multiple-output} (SIMO) communication system in which one or multiple single-antenna UEs transmit pilots to $A$ distributed APs with $M_R$ antennas each, leading to $B=A M_R$ receive antennas in total. 
We consider {orthogonal frequency-division multiplexing} (OFDM) transmission with $W$ occupied subcarriers.
We assume that the AP $a\in\setA=\{1,\dots,A\}$ is at position $\xap{a}\in\reals^3$ in physical space, and that these positions are known.\footnote{It is a reasonable assumption to know the AP positions, as the APs are typically placed according to a carefully-crafted deployment plan.} 

Suppose that we have $N$ pilot transmissions from one UE at (unknown) positions $\bmx^{(n)}\in\opR^3$ at timestamps $t_n$ for $n\in\setN=\{1,\dots,N\}$.
The $n$th transmission from UE position~$\bmx^{(n)}$ enables the AP $a$ to estimate the associated CSI vector $\bmh^{(n,a)}_w\in\opC^{M_R}$ at timestamp~$t_n$ and subcarrier $w$.
By stacking the channel vectors $\{\bmh^{(n,a)}_w\}_{a=1}^A$ from all APs, we can construct a CSI vector $\bmh^{(n)}_w\in\opC^B$ that contains CSI for all $B$ receive antennas.
It is important to note that we do not require the $A$ APs to be perfectly synchronized while acquiring CSI; 
the only requirement is that the CSI estimated at each AP belongs to the same UE transmitting from approximately the same position~$\vecx^{(n)}$ at approximately the same timestamp $t_n$.
Finally, we concatenate the channel vectors from all subcarriers to construct a $B\times W$ CSI matrix associated with the UE at timestamp~$t_n$ as $\bH^{(n)}=[\bmh_1^{(n)},\dots,\bmh_W^{(n)}]$.
The entire CSI database is given by the set of matrices $\{\bH^{(n)}\}_{n\in \setN}$.

\subsection{CSI Feature Extraction}
\label{sec:feature_extraction}

In order to extract large-scale fading properties of the wireless channel~\cite{studer18cc} and to render channel charting resilient to system and hardware impairments (e.g., phase offsets between APs)~\cite{ferrand2021, gonultas22twc}, one transforms the CSI matrices into CSI features. 
We follow the CSI feature design from~\cite{lundpaper,lei19siamese}. 
First, we transform frequency-domain CSI into the delay domain by applying an inverse discrete Fourier transform over the $W$ occupied subcarriers. 
Since most of the received power is typically concentrated in the first few taps, we truncate the delay-domain CSI by taking the first $C$ columns.
In order to render our CSI features resilient to changes in the UEs' transmit power,
we normalize the truncated delay-domain CSI matrix to have unit Frobenius norm and denote the resulting matrix by $\hat\bH^{(n)}\in\opC^{B\times C}$.
Finally, we compute the $D'=BC$ dimensional vector $\vecf^{(n)} = |\mathrm{vec}(\hat \bH^{(n)})|$, which ignores phase shifts (e.g., from APs that are not synchronized in phase).
The entire CSI feature database is given by the set of vectors~$\{\vecf^{(n)}\}_{n\in \setN}$.

\subsection{Channel Charting with Neural Networks}

With the CSI feature database, one can now learn a channel charting function $g_{\boldsymbol\theta}:\opR^{D'}\to\opR^{D}$, which maps a CSI feature vector $\vecf^{(n)}$ to a $D$-dimensional pseudo-position as follows: $\hat\vecx^{(n)} = g_{\boldsymbol\theta}(\vecf^{(n)})$.
The function  $g_{\boldsymbol\theta}$ is implemented as a neural network that is parametrized by the vector~$\boldsymbol\theta$ which includes all of the network's weights and biases. 
While real-word positions are in three-dimensional space with coordinates $(x,y,z)$, the channel chart can be embedded in fewer dimensions if, for example, the $z$-coordinate is fixed across the AP and UE positions. Hence, assuming that $D\leq 3$, we denote the truncated vector consisting of the first $D$ entries of $\vecx^{(n)}$ and $\xap{a}$ as~$\tilde\vecx^{(n)}\in\opR^D$ and~$\tildexap{a}\in\opR^D$, respectively.

The literature describes a variety of  methods to learn neural network-based channel charting functions, such as autoencoders~\cite{studer18cc}, 
Siamese neural networks \cite{lei19siamese}, and neural networks trained with a timestamp-based {triplet loss} \cite{ferrand2021,yassine22,rappaport2021}. 
In what follows, we focus on the triplet-loss channel charting approach from~\cite{ferrand2021}.

\subsection{Triplet-Based Channel Charting}
\label{sec:triplet_cc}

Assuming that the timestamps associated with all CSI features are available, reference~\cite{ferrand2021} proposes to use this side information when comparing pairwise distances in latent space.
To this end, one defines a set of triplets from the set of sample indices $\setN$ as follows:
\begin{align}
\setT = \{(n,c,f) \in \setN^3: 0 < |t_n-t_c| \leq T_{\mathrm{c}} <|t_n-t_f|\},
\end{align}
where $T_{\mathrm{c}}>0$ is a parameter that categorizes the CSI features as close or far in time.
If $t_n$ is closer to $t_c$ than $t_f$, then we would expect the Euclidean distance between the UE positions at timestamp~$t_n$ and~$t_c$ to be smaller than that of $t_f$.
This property can be expressed with the following triplet loss~\cite{ferrand2021}: 
\begin{align} \label{eq:loss_t}
\loss_{\mathrm{t}} = \frac{1}{|\setT|} \! \sum_{(n,c,f)\in\setT} \!\! \big(& \|g_{\boldsymbol\theta}(\vecf^{(n)}) - g_{\boldsymbol\theta}(\vecf^{(c)})\| \notag\\[-0.2cm]
& - \|g_{\boldsymbol\theta}(\vecf^{(n)}) - g_{\boldsymbol\theta}(\vecf^{(f)})\| + M_{\mathrm{t}}\big)^+.
\end{align}
Here, the margin parameter~$M_{\mathrm{t}}\geq0$ enforces $g_{\boldsymbol\theta}(\vecf^{(n)})$ to be at least $M_{\mathrm{t}}$ closer to $g_{\boldsymbol\theta}(\vecf^{(c)})$ than to $g_{\boldsymbol\theta}(\vecf^{(f)})$.

As demonstrated in~\cite{ferrand2021}, one can train a neural network that implements the channel charting function $g_{\boldsymbol\theta}$ by minimizing the loss $\loss_{\mathrm{t}}$; this approach is self-supervised as the loss only utilizes the training dataset without any additional ground-truth labels, such as UE positions.  
We reiterate that the coordinate system of the resulting channel chart is arbitrary, e.g., scaled, rotated, and globally warped.
The method we propose next addresses exactly this limitation.


\section{Channel Charting in Real-World Coordinates}
\label{sec:ccinreal}

We now propose a novel loss that enables one to directly learn channel charts in real-world coordinates.
Our method only uses the known AP positions in a multipoint scenario---\emph{no} geometric models, accurate time synchronization among APs, or  CSI samples labeled with UE positions are required.

\subsection{Bilateration Loss}
\label{sec:loss_b}

Assuming that a UE's pilot signal is received at multiple APs, there is one clue about the position of the UE that our CSI features contains: the relative receive power at each AP. 
As one would expect, an AP with higher relative power tends to be closer to the UE under line-of-sight conditions.
Let~$\hat\bH^{(n,a)} \in \opC^{M_R\times C}$ denote the submatrix of~$\hat\bH^{(n)}$ which consists of the $M_R$ rows corresponding to AP $a$.
We now compute the relative power of the channel between the UE and the $a$th AP at timestamp~$t_n$ as $P^{(n,a)} = 20\log_{10} ( \| \hat\bH^{(n,a)}\|_F )$, $a\in\setA$,
and define a set of AP pairs 
\begin{align} \label{eq:setP}
\setP^{(n)} = \big\{(a_c,a_f) \in \setA^2  : P^{(n,a_c)} > P^{(n,a_f)} + M_{\mathrm{p}}\big\}.
\end{align}
The set $\setP^{(n)}$ generates pairs of AP indices $(a_c,a_f)$, where the two selected APs act as close and far reference points for the $n$th UE position. 
Here, the margin parameter $M_{\mathrm{p}}\geq 0$ enforces the set $\setP^{(n)}$ to only include AP pairs whose channel powers differ by at least $M_{\mathrm{p}}$.
The assumption that the UE $n$ should be closer to AP $a_c$ than AP $a_f$ as determined by the set $\setP^{(n)}$ in \fref{eq:setP} leads to the following \textit{bilateration loss}: 
\begin{align} \label{eq:loss_b}
\loss_{\mathrm{b}}= & \frac{1}{|\setN||\setP^{(n)}|} \sum_{n\in\setN} \sum_{(a_c,a_f)\in \setP^{(n)}} \notag \\
& \big( \|g_{\boldsymbol\theta}(\vecf^{(n)}) \!-\! \tildexap{a_c}\|  \!-\! \|g_{\boldsymbol\theta}(\vecf^{(n)}) \!-\! \tildexap{a_f}\|+ M_{\mathrm{b}}\big)^+\!\!.
\end{align}
Here, the margin $M_{\mathrm{b}}\geq0$ enforces $g_\theta(\vecf^{(n)})$ to be at least $M_{\mathrm{b}}$ closer to AP $a_c$  than AP $a_f$.
We emphasize that the loss $\loss_{\mathrm{b}}$ requires no assumption of how far each AP should be from the UE based on their channel power; we merely deduce relative distances to the two APs. Moreover,
while the underlying assumption on the power-distance relation between UE and APs may not always hold true, we seek no such guarantee---we support this claim with a concrete example in \fref{sec:power_vs_distance}.
One can train a neural network that implements the channel charting function~$g_{\boldsymbol\theta}$ by minimizing the loss $\loss_{\mathrm{b}}$ in~\fref{eq:loss_b}; 
this approach is weakly-supervised as the loss incorporates only partial information on the ground-truth labels (i.e., AP positions).

We note that a loss based on received power was also proposed in~\cite{karmanov21wicluster}; this loss, however, compares the power of two CSI vectors corresponding to the channel between one AP and the UE at two timestamps $t_n$ and $t_f$  
in order to deduce whether $\vecx^{(n)}$ or $\vecx^{(f)}$ is closer to the AP. 
This loss requires the UEs to have fixed transmit power over time and does not leverage the fact that the UEs are simultaneously transmitting to multiple APs.
In contrast, our bilateration loss in \fref{eq:loss_b} takes each CSI feature into account individually and compares the channel power between a pair of APs; this means that our loss does not rely on timestamps and does not require the transmit power to stay constant between samples.
Thus, our bilateration loss could also be used in scenarios in which timestamps are irrelevant, e.g., a multiuser scenario with stationary UEs.

\subsection{Combined Triplet and Bilateration Loss}
\label{sec:loss_combined}

In order to improve channel charting, we propose to combine the two loss functions $\loss_{\mathrm{t}}$ {in~\fref{eq:loss_t}} and $\loss_{\mathrm{b}}$ {in~\fref{eq:loss_b}}, since $\loss_{\mathrm{t}}$ helps in preserving local neighborhood relations whereas $\loss_{\mathrm{b}}$ helps to anchor the channel chart in real-world coordinates; 
moreover, false triplets (i.e., triplets for which the anchor point is closer in space to the point whose timestamp is further apart) in $\loss_{\mathrm{t}}$ and false AP pairs (see~\fref{sec:power_vs_distance} for a detailed definition) in~$\loss_{\mathrm{b}}$
can be counterbalanced. 
We define the combined triplet and bileration loss as follows:
\begin{align} \label{eq:loss_combined}
\loss_{\mathrm{t},\lambda \mathrm{b}} = \loss_{\mathrm{t}} + \lambda\loss_{\mathrm{b}}. 
\end{align}
Here, $\lambda>0$ is a parameter that balances the two losses; the choice of this parameter is briefly discussed in \fref{sec:methods}. 

One can train a neural network that implements the channel charting function $g_{\boldsymbol\theta}$ by minimizing the loss $\loss_{\mathrm{t},\lambda \mathrm{b}}$; this approach is a hybrid between self- and weakly-supervised learning due to the reasons mentioned in \fref{sec:triplet_cc} and \fref{sec:loss_b}, respectively.

\section{Results}
\label{sec:results}

We now present results for the proposed channel charting methods and comparisons to various baselines.

\subsection{Simulation Scenario}
\label{sec:simulation_scenario}

We consider an urban scenario in which multiple APs are placed around a rectangular area of size $83\,\text{m} \times 122\,\text{m}$; see \fref{fig:channel_charts}\,(a) for an illustration of the UE and AP positions. We use sub-6-GHz scenario with channel vectors from Remcom's Wireless InSite ray-tracing software \cite{remcom}. 
\fref{tbl:simulation_parameters} summarizes the simulation parameters.
Each CSI matrix $\bH^{(n)}$, $n\in\setN$ is of dimension $32\times1200$; each CSI feature vector is of dimension $D'=32\times8=256$, where we use the $C=8$ first taps and the feature extraction pipeline from \fref{sec:feature_extraction}. 
Since the heights of all APs are the same and the height of all UEs is also fixed, we perform channel charting in $D=2$ dimensions.

\begin{table}[tp]
\centering
\caption{Summary of Simulation Parameters}\label{tbl:simulation_parameters}
    \begin{tabular}{@{}ll@{}}
        \toprule
        Parameter & Value or Type \\
        \midrule
        Number of APs & $A=8$\\
        Number of antennas per AP & $M_R=4$\\
        Number of UE positions & $15\,912$\\
        Spacing between UE positions & $0.8$\,m \\
        AP antenna height & $10$\,m\\
        AP antenna spacing & Half-wavelength\\
        UE antenna height & $1.5$\,m\\
        AP and UE antenna type & Omnidirectional\\
        Carrier frequency & $1.9$\,GHz\\
        Bandwidth & $20$\,MHz\\
        Number of used subcarriers & $W=1200$ \\
        Received SNR per AP & $-9.7$\,dB to $33.8$\,dB \\
        \bottomrule
    \end{tabular}		
\end{table}

\subsection{Is Received Power Inversely Proportional to Distance?}
\label{sec:power_vs_distance}

\begin{figure}[tp]
\centering
\includegraphics[width=0.98\columnwidth]{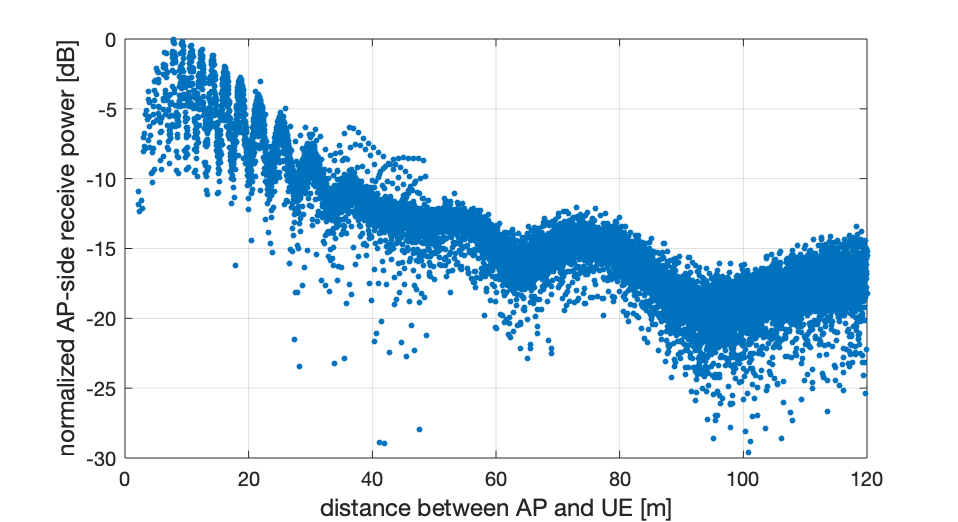}
\vspace{-0.1cm}
\caption{Receive power (in decibels) at one AP dependent on the AP-to-UE distance (in meters) in the xy-plane. The receive power values are normalized to the range $(-\infty,0]$\,dB.} 
\label{fig:pow_vs_distance}
\end{figure}

Before we show channel charting results, we first provide an example of the power-distance relation that is exploited by our bilateration loss in \fref{eq:loss_b}. 
\fref{fig:pow_vs_distance} shows the (normalized) receive power at one AP depending on the distance to an UE in the xy-plane; we ignore the distance in the z-axis since the heights are fixed. 
Evidently, the power generally decreases with distance, but small-scale fading causes the power to fluctuate quite substantially among the different UE positions which are at approximately the same distance to the AP.
Furthermore, the received power increases\footnote{This behavior can be attributed to the antenna radiation pattern, which causes the receive power to reduce if a UE approaches the area below an AP.} with distance until about $10$\,m, which negatively affects approaches that rely on the power being the highest near the AP, e.g., the method in~\cite{pihljasalo20}. 

Since a smaller difference in power is more likely to be caused by small-scale fading (and not by distance), we would be less confident in deducing which AP might be closer. Hence, 
observing that the AP-side receive power is not perfectly inversely proportional to distance motivates the margin  $M_{\mathrm{p}}$ in \fref{eq:setP}.
By setting $M_{\mathrm{p}}>0$, we can avoid some false AP pairs, i.e., pairs $(a_c,a_f) \in \setP^{(n)}$ for which $\|\tilde\vecx^{(n)}- \tildexap{a_c}\| > \|\tilde\vecx^{(n)}- \tildexap{a_f}\| $.
For our CSI dataset, margins $M_{\mathrm{p}}$ of $0$, $3$, and $6$ result in false AP pair ratios of  $19.5$\%, $9.9$\%, and $4.9$\% for all UEs, 
while the average number of AP pairs in the set $\setA^{(n)}$ given by $\frac{1}{N}\sum_{n=1}^N |\setA^{(n)}|$ is $28$, $17.9$, and $10.1$, respectively.
Clearly, there exists a trade-off in choosing~$M_{\mathrm{p}}$:
Increasing $M_{\mathrm{p}}$ leads to fewer \emph{false} AP pairs at the cost of fewer AP pairs,
which yields more accurate but fewer reference points for weakly-supervised learning (and vice versa).

\subsection{Learning the Channel Charting Function}
\label{sec:nn_structure}

{Following the neural network architecture in \cite{ferrand2021},} we use a six-layer fully-connected neural network for the channel charting function $g_{\boldsymbol\theta}$. The neural network has $\{256, 128, 64, 32, 16, 2\}$ activations per layer; all layers except for the last one use ReLU activations, whereas the last layer uses linear activations.
We initialize the weights with Glorot~\cite{glorot} and use the Adam optimizer to learn the channel charting function $g_{\boldsymbol\theta}$. 

The CSI dataset described in \fref{sec:simulation_scenario} is created for UE positions on a $104\times153$ rectangular grid with a total of $15\,912$ CSI samples.
We split this CSI sample set randomly into training and test sets of $12\,000$ and $3912$ samples, respectively.
For training with the triplet loss, we timestamp the CSI dataset by imitating the following trajectory in the rectangular area: 
The UE first moves on the outermost (largest) rectangle in the area on the grid, then takes one step inwards, walks on the rectangle whose edges are now $1.6$\,m shorter than the first rectangle, then takes one step inwards, {and so on}, until the innermost (smallest) rectangle is completed and all points in the dataset are covered.

\subsection{Proposed Methods and Baselines}
\label{sec:methods}

In order to compare our proposed methods to baselines, we also consider two positioning methods and state-of-the-art channel charting.
All of these methods (including ours) use the same neural network architecture (cf.~\fref{sec:nn_structure}) but are trained with different loss functions, which we detail next. 

\subsubsection*{Proposed\,\,1 (P1)}
\label{sec:bilateration_loss}

For our first proposed method, 
we extract the relative power of the (normalized) CSI feature vector to each AP as described in \fref{sec:loss_b}.
We train the neural network only using the bilateration loss  $\loss_{\mathrm{b}}$ defined in \fref{eq:loss_b} with parameters $M_{\mathrm{p}}=3$ and $M_{\mathrm{b}}=15$. We reiterate that, due to the reasons mentioned in \fref{sec:loss_b}, this approach is weakly-supervised.
In addition, this method also demonstrates the use of large $\lambda$ values in \fref{eq:loss_combined} as $\loss_{\mathrm{b}} = \lim_{\lambda\to \infty}  {\lambda}^{-1}\loss_{\mathrm{t},\lambda \mathrm{b}}$.

\subsubsection*{Proposed\,\,2 (P2)}
\label{sec:combined_triplet_bilateration_loss}

For our second proposed method, we assume that timestamps are available for all CSI features in the training set; we also extract the relative power of the (normalized) CSI feature vector to each AP as described in \fref{sec:loss_b}.
We train the neural network using the combined triplet and bilateration loss  $\loss_{\mathrm{t},\mathrm{b}}$ defined in~\fref{eq:loss_combined} with parameters $T_{\mathrm{c}}=20$, $M_{\mathrm{p}}=3$, and $M_{\mathrm{t}}=M_{\mathrm{b}}=15$.
We reiterate that due to the reasons mentioned in \fref{sec:triplet_cc} and \fref{sec:loss_b} regarding each component of the loss $\loss_{\mathrm{t}}$ and $\loss_{\mathrm{b}}$, this approach is a hybrid between self- and weakly-supervised learning.

\subsubsection*{Baseline\,\,1 (B1)}
\label{sec:mse_loss}

In order to compare our methods to supervised positioning that processes CSI features with neural networks (see, e.g.,~\cite{ferrand2020globecom,gonultas22twc} 
and the references therein), this baseline assumes that ground truth positions are available for all samples in the training set. Therefore, we train this neural network with the mean squared error (MSE) loss:
\begin{align}\label{eq:loss_mse}
    \loss_{\MSE} = \frac{1}{|\setN|} \sum_{n\in\setN}  \|g_{\boldsymbol\theta}(\vecf^{(n)}) - \tilde\vecx^{(n)}\|^2.
\end{align}
Since the loss $\loss_{\MSE}$ requires ground-truth labels (i.e., UE positions) for all training samples, this approach relies on supervised learning.
This method serves as a baseline to show the best-possible positioning performance achievable with the neural network architecture described in~\fref{sec:nn_structure}. 

\subsubsection*{Baseline\,\,2 (B2)}
\label{sec:triplet_loss}

In order to compare our methods to the {state of the art} in channel charting, 
this baseline assumes that the timestamps are available for all CSI features in the training set.
We train the neural network using the triplet loss  $\loss_{\mathrm{t}}$ defined in \fref{eq:loss_t} with parameters $T_{\mathrm{c}}=20$ and $M_{\mathrm{t}}=1$. 
We reiterate that this approach is self-supervised due to the reasons mentioned in \fref{sec:triplet_cc}.
Furthermore, this baseline also demonstrates the result for using small values of $\lambda$ in \fref{eq:loss_combined} as $\loss_{\mathrm{t}} = \lim_{\lambda\to 0} \loss_{\mathrm{t},\lambda \mathrm{b}}$.
Note that the resulting channel charts are typically of high quality but embedded in arbitrary coordinates.

\subsubsection*{Baseline\,\,3 (B3)}
\label{sec:combined_triplet_mse_loss}

In order to compare our methods to semi-supervised positioning techniques \cite{lei19siamese,deng21networkside,karmanov21wicluster,zhang21globecom,deng21tsne,penghzipaper},
this baseline assumes that while the timestamps are available for all samples in the training set, UE ground-truth position information is available for a small subset $\setN'$ of the training set, i.e., $|\setN'| = 0.05N$. 
Hence, the MSE loss can be calculated only for a subset of the available samples.
Although the MSE loss is the same as in \fref{eq:loss_mse},
we denote the loss in this case by $\loss_{\widetilde{\MSE}}$ to distinguish it from the loss taken over all training samples in the fully-supervised baseline B1.
We train the neural network using the combined triplet and $\widetilde{\text{MSE}}$ loss function defined as
\begin{align}
    \loss_{\mathrm{t},\widetilde{\MSE}}  = \loss_{\mathrm{t}}+\loss_{\widetilde{\MSE}} \label{eq:loss_tMSE}
\end{align}
with  parameters $T_{\mathrm{c}}=20$ and $M_{\mathrm{t}}=15$.
Since the loss $\loss_{\mathrm{t},\widetilde{\MSE}}$ utilizes a subset of the UE ground-truth positions in addition to the triplet loss, this approach is semi-supervised.
This baseline positioning method showcases whether our proposed methods can achieve the performance of semi-supervised techniques without the need for UE ground-truth position information.

\subsection{Performance Metrics}
\label{sec:perf_metrics}

We evaluate the performance of the proposed methods and baselines using {six} different metrics.
The first four metrics measure the quality of the latent space; we refer to~\cite{altous22asilomar} for the details.
The latter two metrics measure positioning accuracy:
 (i) \emph{Trustworthiness (TW)} penalizes neighborhood relationships in latent space that are not present in the real-world positions~\cite{vathy2013}; TW values are in $[0,1]$ with optimal value~$1$. 
 (ii) \emph{Continuity (CT)} measures how well neighborhood relationships between the real-world positions are preserved in latent space~\cite{vathy2013}; CT values are in $[0,1]$ with optimal value~$1$.
 (iii) \emph{Kruskal stress (KS)} measures the dissimilarity between pairwise distances in the real-world positions and pairwise distances in latent space~\cite{kruskal1964multidimensional}; KS values are in $[0,1]$ with optimal value~$0$.
 (iv) \emph{Rajski distance (RD)} measures the difference between the mutual information and joint entropy of the distribution of pairwise distances in the real-world positions and latent space~\cite{rajski1961}; RD values are in $[0,1]$ with optimal value~$0$. 
 (v) \emph{Mean distance error} measures the average  error in the UE position estimates in  Euclidean norm over all UE positions; this metric is nonnegative with optimal value~$0$.
 (vi) \emph{95th percentile distance error} measures the 95th percentile of the error in the UE position estimates in   Euclidean norm over all UE positions; this metric is nonnegative with optimal value~$0$. 

\subsection{Results and Comparison}
\label{sec:perf_comparison}

We now compare the proposed methods with the baselines discussed in \fref{sec:methods} using the metrics from \fref{sec:perf_metrics}; all of the channel charts, positioning plots, and evaluated metrics are based on the test set of $3912$ CSI samples mentioned in \fref{sec:nn_structure}. 
\fref{fig:channel_charts} visually compares the ground-truth positions as well as channel charts in arbitrary and real-world coordinates; \fref{tbl:result_table} summarizes the types of supervision and the associated performance metrics. 
Note that the two distance error metrics are irrelevant for baseline B2 as it produces channel charts in arbitrary coordinates; hence, we do not report these metrics for baseline B2 in \fref{tbl:result_table}.

Figures~\ref{fig:channel_charts}\hspace{0.02cm}(b-f) show that all of the considered methods preserve local geometry. 
As expected, the baselines B1 and B3 result in the most visually pleasing results, as they rely on ground-truth UE position information.
In contrast, B2 provides a high-quality channel chart but in arbitrary coordinates and where global geometry is distorted. 
The proposed method P1 provides a channel chart in real-world coordinates, but of slightly worse quality than the triplet-loss-based channel charting baseline B2.
The proposed method P2 combines the best of both worlds: the quality of the channel chart from B2 and representation in real-world coordinates from P1.

\fref{tbl:result_table} reveals that the supervised positioning baseline B1 achieves the best performance in all considered metrics.
We observe that, while the proposed method P1 has the advantages of (i) generating a channel chart in real-world coordinates 
and (ii) not requiring \emph{any} ground-truth AP or UE position labels, it is outperformed by all the other methods in terms of all latent space quality metrics;
we attribute the lower quality in P1's channel chart to the false AP pairs  mentioned in \fref{sec:power_vs_distance}.
We also observe that the proposed method P2 outperforms B2, B3, and P1 in all latent space quality metrics, while its mean and 95th percentile positioning errors are only slightly higher than those of B3 by $0.44$\,m and $1.72$\,m, respectively.
Furthermore, the mean and 95th percentile positioning errors of P2 are only $5.42$\,m and $12.31$\,m higher than those of the fully-supervised positioning baseline B1.
Here, we reiterate that the proposed method P2 does not require any ground-truth UE position information as opposed to B1 and B3; hence, it is remarkable that P2 achieves similar positioning performance to the {supervised positioning baseline} B1 and approximately the same positioning performance as {the semi-supervised baseline} B3, {without the need for ground-truth UE position information}. 

\newcommand{\figsize}{0.3}
\begin{figure*}
    \centering
    \subfigure[Ground-truth UE positions]
    {
    \includegraphics[width=\figsize\textwidth]{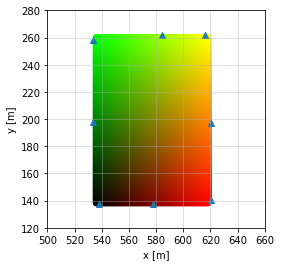}
    }\label{fig:gt_pos}
    \subfigure[P1 (weakly supervised)]  
	{
    \includegraphics[width=\figsize\textwidth]{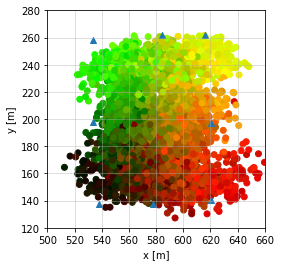}
    }\label{fig:onlyaploss}
    \subfigure[P2 (hybrid self- and weakly-supervised)]
	{
    \includegraphics[width=\figsize\textwidth]{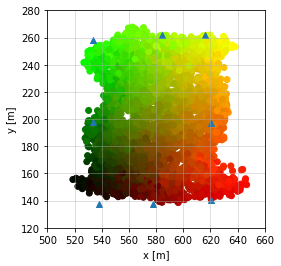}
    }\label{fig:bothlosses}
    
    \vspace{0.1cm}
    
     \subfigure[B1 (supervised)]
	{
    \includegraphics[width=\figsize\textwidth]{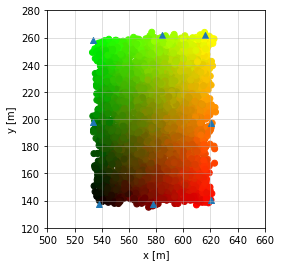}
    }\label{fig:sup}
    \subfigure[B2 (self-supervised)]
	{
    \includegraphics[width=\figsize\textwidth]{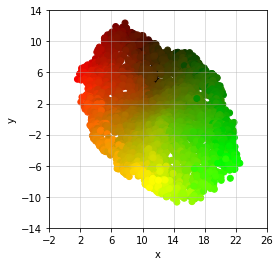}
    }
    \subfigure[B3 (semi-supervised)] 
	{
    \includegraphics[width=\figsize\textwidth]{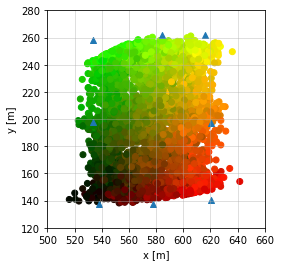}
    }\label{fig:semisup}
    \caption{(a) UE ground-truth positions (green-to-red gradient-colored area), AP positions (blue triangles), {and the channel charts or positioning estimates (b-f)} for the proposed (P) and baseline (B) methods. Since the output of baseline B2 is in arbitrary coordinates, the AP positions are not shown in (e). The proposed method P2 achieves comparable results in real-world coordinates as the semi-supervised baseline B3 but without requiring known UE positions during training.} 
    \label{fig:channel_charts}   
\end{figure*}

\begin{table*}
\centering
\resizebox{0.92\textwidth}{!}{
\begin{minipage}[c]{0.99 \textwidth}
    \centering
    \caption{Channel charting and positioning performance comparison.}
    \label{tbl:result_table}		 
    \begin{tabular}{@{}lllccccccc@{}}
        \toprule
        && & &\multicolumn{4}{c}{Latent space quality metrics} & \multicolumn{2}{c}{Positioning error [m]} \\
        \cmidrule(lr){5-8} \cmidrule(lr){9-10}  
        Method & {Loss function to learn $g_{\theta}$} & Learning type & Figure & TW$\,\uparrow$ & CT$\,\uparrow$ & KS$\,\downarrow$ & RD$\,\downarrow$  &  Mean$\,\downarrow$ & 95th percentile$\,\downarrow$ \\
        \midrule
        P1 & {Eq.\,\ref{eq:loss_b}, bilateration} & Weakly-supervised & {2\,(b)}& 0.918 & 0.922 & 0.26 & 0.88 & 15.02 & 32.05  \\ 
        P2 & {Eq.\,\ref{eq:loss_combined}, bilateration and triplet} & Hybrid self- and weakly-supervised & {2\,(c)}& 0.990 & 0.990 & 0.12 & 0.71 & 7.64 & 16.82 \\ 
        \midrule
        B1 & {Eq.\,\ref{eq:loss_mse}, MSE} & Fully-supervised  & {2\,(d)}& 0.998 & 0.998 & 0.04 & 0.45 & 2.22 & 4.51  \\ 
        B2 & {Eq.\,\ref{eq:loss_t}, triplet} & Self-supervised & {2\,(e)}& 0.991 & 0.992 & 0.17 & 0.79 & -- & -- \\ 
        \vspace{-0.4cm}\\
        B3 & {Eq.\,\ref{eq:loss_tMSE}, triplet and $\widetilde{\text{MSE}}$} & Semi-supervised & {2\,(f)}& 0.980 & 0.980 & 0.13 & 0.72 & 7.20 & 15.10   \\ 
        \bottomrule
    \end{tabular}		
\end{minipage}
}\vspace{0.2cm}
\end{table*}


\section{Conclusions and Future Work}
\label{sec:conclusions}

We have proposed a novel bilateration loss that enables weakly-supervised channel charting in real-world coordinates. 
This loss utilizes the known AP locations in a multipoint scenario by comparing the received power at two APs and placing the UE closer to the one with higher power. 
Using channel vectors from a commercial ray-tracer, we have demonstrated that combining the bilateration loss with the triplet loss from~\cite{ferrand2021} is sufficient to generate high-quality channel charts in real-world coordinates, without the need of geometric models, accurate AP synchronization, or ground-truth UE positions. 

There are many avenues for future work. 
We expect that loss functions that compare AP angle-of-arrival information besides receive power would further improve the quality of the learned channel charts. 
Extending our approach to indoor scenarios and measured channels under non-line-of-sight conditions, which requires AP selection, is part of ongoing work.

\bibliographystyle{IEEEtran}

\balance

\bibliography{bib/IEEEabrv,bib/confs-jrnls,bib/publishers,bib/studer,bib/vipbib, bib/cc_bib, bib/emres_bib} 

\balance
	
\end{document}